\documentstyle[12pt]{report}

\renewcommand{\abstract}[1]{{\footnotesize\noindent{\bf Abstract} #1 \\}}
\renewcommand{\author}[1]{\subsubsection*{#1}}
\newcommand{\address}[1]{\subsubsection*{\it#1}}

\begin{document}

\chapter*{Gamma-Ray Bursts and TeV quantum gravity}

\author{ D.I.  Kosenko$^{1,2}$  K.A. Postnov$^{1,2}$}

%\titleauthor{}{}

\address{$^1$ Faculty of Physics, Moscow State University, 119899 Moscow, Russia\\
$^2$ Sternberg Astronomical Institute, 119899 Moscow, Russia}

\abstract{
Limitations on the multi-dimensional TeV-scale quantum gravity model by
Arkani-Hamed, Dimopoulos and Dvali (1998) and on model by Dvali,
Gabadadze, Porrati (2000) are obtained from an analysis
of gamma-ray bursts at cosmological distances (relativistic fireball
model). }

\section{Extra dimensions and their manifestations}

Recently an idea of large extra dimensions become very popular due to
benefits it provides in solving of hierarchy problem of particle physics 
(the fundamental Plank mass scale 
$\sim 10^{18}$ GeV relevant for gravitation is much larger
than electroweak scale $\sim 1$ TeV (e.g. \cite{Ru_UFN} for a review).
These theories assume that the Standard Model particles are localized
in a (3+1)-dimensional "brane" embedded in a compactified space (bulk)
with large (or infinitely large) extra dimensions. In these scenarios,
the fundamental gravity scale is no more the conventional Planck mass, 
which determines the observable weakness of the Newton gravitational 
constant $G_N$. The latter turns out to be defined by the quantum
gravity scale $M_*$ of the corresponding theory.

In such a frame, one of the phenomenological manifestations of the
existence of large (or infinite) extra dimension(s) is an additional
cooling of hot plasma due to emission of Kaluza-Klein massive gravitons
into the bulk (ADD model, \cite{ADD}) or excitation of stringy Regge
states (DGP model, \cite{DGP}): which are allowed to propagate freely
in the bulk and carry away energy from the brane. Though this coupling
is suppressed by 4-dimensional Planck mass scale, at high enough
energies it may become noticeable, and it is conceivable to observe
events with energy missing. 

\section{Constraints on the models parameters}

There are number of restrictions on fundamental gravity
scale $M_*$ of models with extra dimensions from high energy phenomena
in astrophysics and from the processes in the early Universe
\cite{ADD,Ru_UFN}. The basic constraints come from SN1987a cooling,
BBN, etc... (see \cite{ADD, Ru_UFN, HR, HR_NS, WD} for more detail). 
Here we consider constraints the very existence of cosmological gamma-ray
bursts (GRB) imposes on some modern theories of gravity. As an example, we
examine the ADD theory of multi-dimensional gravity with quantum gravity
scale at TeV energies (\cite{ADD}), and more recent DGP 5D-gravity of
infinite-volume flat extra space with $10^{-3}$ eV quantum gravity scale
(\cite{DGP,DGKN}).

The most viable model of GRB is the relativistic fireball model, which
is apparently confirmed by the bulk of GRB studies in a wide range of
wavelengths from radio to gamma-rays (see Dr.~M.Vietri's lecture, this
volume; \cite{P1, P2}). In a GRB, a huge energy ($10^{51}-10^{53}$
ergs) in gamma-rays ($E_\gamma\sim 100$ keV -- 10 MeV) is released in a
short time (typically observed $\Delta t_\gamma \sim 10-100$ s). This
energy, liberated in a small region $\sim 10^6-10^7$ cm in size,
creates a photon-lepton "fireball" with a very high 
energy densities. For the characteristic energy $E_{53}=\Delta
E_\gamma/10^{53}$ erg and the initial size $r_6=r_0/10^6$ cm, the
initial temperature of the optically thick fireball is $
T_f\simeq 116 (\hbox{MeV})E_{53}^{1/4}r_6^{-3/4}$.
The photon number density (as well as of relativistic leptons) is 
$n_\gamma\simeq 4.3\times 10^{37}(\hbox{cm}^{-3})(T/100\,\hbox{MeV})^3$
and diverse photon-photon and photon-lepton processes intensively
occur. Thus the GRB fireballs can be potentially useful to test
high-energy physics at MeV scales. 

In the ADD scenario, the 4D Planck mass is related to the
compactification radius $r_n$ and fundamental gravity scale $M_*$
as $M_P\sim r_n^n M_*^{n+2}$, where $n$ is the number of extra
dimensions. The emission of KK-gravitons in the bulk in
photon-photon interactions (relevant to GRB fireballs) has a cross
section (\cite{ADD})
$$
\sigma_{\gamma\gamma}\sim \frac{1}{16\pi}\left(\frac{T}{M_*}\right)^n
\frac{1}{M_*^2}
$$
i.e. the KK-luminosity becomes
$$
(dE/dt)_{KK}\sim n_\gamma^2\sigma_{\gamma\gamma}c\epsilon_{KK}
\propto T_f^{7+n}/M_*^{2+n}
$$
(here $\epsilon_{KK}\sim 2.7 T_f$ is the typical KK-graviton energy).

If the emission of KK-gravitons effectively cools the
fireball before its initial thermal energy is converted into
the kinetic energy of the baryons, the required high
Lorentz-factors can not be attained, and no GRB with
the observed properties can be produced. This implies that
the emission of KK-gravitons in the fireball meets
the condition $r_0/c < \Delta E_\gamma/(dE/dt)_{KK}$.
Putting all quantities together, we arrive at the following constraints:
$$
n=2:\quad M_*>2(\hbox{TeV})E_{53}^{5/16}r_6^{-11/16}\,,
$$
$$
n=3:\quad M_*>0.25(\hbox{TeV})E_{53}^{3/10}r_6^{-7/10}\,.
$$
These are weaker than the limits inferred from SN1987a neutrino burst
($M_*>30$ TeV for n=2) and from cosmological considerations
\cite{ADD, HR}.

More interesting is the case of DGP model. In this framework, the
weakness of an observable gravity is explained by the high cut-off
of the Standard Model $M_{SM}$ localized on the brane. In contrast
to the ADD model, the large value of the observable $M_P$ is
determined by $M_{SM}\gg M_*$ rather than $M_*$. Now the emission
of massive KK-gravitons into the bulk is strongly suppressed.
Instead, the possibility to produce an exponentially large number
of Regge states at very low energy appears. At $T\ll 1$ TeV, the
total rate of the production of stringy Regge states is determined
by the 2-d mass level and is (\cite{DGKN})
$$
\Gamma_2\sim E\frac{E^4}{M_*^2M_P^2}
$$
(here the mean energy particle $E\sim 2.7 T$).
This gives rise to the total Regge state emission rate in
the GRB fireball $(dE/dt)_R\approx 10^{55} (\hbox{erg/s}) 
E_{53}^{9/4}r_6^{-15/4}$ 
and the fireball acceleration constraints would be
$ M_*>0.5\, (\hbox{eV}) E_{53}^{5/8}r_6^{-11/8} $.
This is by about two orders of magnitude higher than original
lower bound $10^{-3}$ eV discussed in \cite{DGKN}. If this limit
is true, deviations from the Newton gravity are expected at
distances smaller than $r<1/M_*\simeq 10^{-3}$ mm.

{\it Acknowledgements}. The work was partially supported by RFBR
grants 99-02-16205, 00-02-17884a, 01-15-99310 and 00-02-17164.

\end{document}